\documentclass[preprint,prd,nofootinbib,tightenlines,amsmath]{revtex4}
\usepackage{graphicx}
\usepackage{bm}
\usepackage{subfigure}

\usepackage{color}
\usepackage{amsmath,amssymb}

\begin{document}
\baselineskip=15pt \parskip=5pt

\vspace*{3em}

\title{$\mu - e$ Conversion With Four Generations}

\author{N.G. Deshpande$^1$, T. Enkhbat$^2$, T. Fukuyama$^3$, X.-G. He$^{2,4}$,  L.-H. Tsai$^2$, K. Tsumura$^2$}

\affiliation
{$^1$Institute for Theoretical Sciences, University of Oregon, Eugene, OR97401, USA\\
$^2$Department of Physics and  Center for Theoretical Sciences, \\
National Taiwan University, Taipei, Taiwan\\
$^3$Department of Physics, Ritsumeikan University, Kusatsu, Japan\\
$^4$INPAC, Department of Physics, Shanghai Jiao Tong University, Shanghai, China
}

\date{\today $\vphantom{\bigg|_{\bigg|}^|}$}

\begin{abstract}
We study $\mu - e$ conversion with sequential four generations. A large mass for the fourth generation neutrino can
enhance the conversion rate by orders of magnitude. We compare constraints obtained from $\mu - e$ conversion using experimental bounds on various nuclei with
those from $\mu \to e \gamma$ and $\mu \to e\bar e e$. We find that the current bound from $\mu - e$ conversion with Au puts the most
stringent constraint in this model. The relevant flavor changing
parameter $\lambda_{\mu e} = V^*_{\mu 4}V_{e4}^{}$ is constrained to be less than $1.6\times 10^{-5}$ for the fourth generation neutrino mass larger than 100 GeV.
Implications for future $\mu -e$ conversion, $\mu \to e\gamma$ and $\mu \to e\bar e e$ experiments are discussed.
\end{abstract}

\maketitle

\section{Introduction}

The standard model (SM) of electroweak and strong interactions based on the gauge group $SU(3)_C\times SU(2)_L\times U(1)_Y$
has been rigorously tested in many ways\cite{Ref:PDG}, in different sectors involving gauge bosons, quarks and leptons. The flavor physics in the quark
sector including CP violation is well described by the Cabibbo-Kobayashi-Maskawa (CKM) mixing matrix\cite{Ref:Cabibbo,Ref:KM} in the charged current interactions of the $W$ boson with three generations of quarks.
Corresponding interactions in the leptonic sector are not on the same footing. In the simplest version of the SM, there are no right handed neutrinos, and the left handed neutrinos are massless. This theory conserves lepton flavor of each family separately, and no flavor changing neutral current (FCNC) processes are allowed in the leptonic sector. However, the observation of neutrino oscillations requires both a non-zero neutrino mass and lepton flavor violation (LFV). The SM has to be extended to give neutrino masses and explain their mixing through a mixing matrix in the charged currents analogous to the CKM matrix in the quark sector, the Pontecorvo-Maki-Nakagawa-Sakata (PMNS) mixing matrix\cite{Ref:Pontecorvo,Ref:MNS}. There are many ways to extend the SM to have neutrino masses and mixing, for example, introduction of right-handed neutrinos to have Dirac mass, or to invoke seesaw mechanism to have Majorana mass, or to have loop induced masses. The existence of mixing among different lepton generations will induce FCNC interactions with LFV.
To further understand the properties of the underlying theory, one should not only study whether a model can produce the correct neutrino masses and their mixing, it is also necessary to study possible implications for other FCNC interaction and confront them with existing and future experimental data. A particularly interesting process is $\mu - e$ conversion.

$\mu - e$ conversion has been studied both theoretically and experimentally~\cite{Ref:Mu2Ecv3,Ref:Mu2Ecv4,Ref:Mu2Ecv1,Ref:Mu2Ecv2}. The quantity measuring the strength of this leptonic FCNC process is $B^A_{\mu \to e} = \Gamma^A_{conv}/\Gamma^A_{capt} =
\Gamma(\mu^- + A(N,Z) \to e^- +A(N,Z))/\Gamma(\mu^- + A(N,Z) \to \nu_\mu + A(N+1, Z-1)$. Experimental bounds  on $\mu - e$ conversion for several nuclei have been obtained as $B^{\text{Au}}_{\mu \to e} < 7\times 10^{-13}$\cite{Ref:Mu2E-Au}, $B^{\text{S}}_{\mu \to e} < 7\times 10^{-11}$\cite{Ref:Mu2E-S}, $B^{\text{Ti}}_{\mu \to e} < 4.3\times 10^{-12}$\cite{Ref:Mu2E-Ti} and $B^{\text{Pb}}_{\mu \to e} < 4.6\times 10^{-11}$\cite{Ref:Mu2E-Pb} with the 90\% c.f. level. These limits can give important information for leptonic FCNC interactions.

There are many different ways to extend the SM to have neutrino masses and mixing, and therefore leptonic FCNC interactions. In this work, we study implications on $\mu - e$ conversion, in one of the simplest extension of the SM by introducing right-handed neutrinos to have Dirac masses. 
In this model, $\mu - e$ conversion cannot occur at tree level. At one loop level, it can be induced by the photon and $Z$ penguin, and box diagrams. Since the Glashow-Iliopoulos-Maiani(GIM) mechanism is in effect, all leptonic FCNC processes involving charged leptons are proportional to $m^2_{\nu_i}/m^2_W$, here $m_\nu$ is the neutrino mass and $m_W$ is the $W$ boson mass.
In the SM with three generations (SM3), due to the very tiny neutrino masses ($\lesssim 0.2$ eV), all LFV rates are extremely small. If the model is further extended to have a fourth generation (SM4), and if the fourth generation neutrino mass $m_{\nu_4}$ is large enough, the factor $m^2_{\nu_i}/m^2_W$ can even become an enhancement factor, and therefore can have observable leptonic FCNC effects.
In other words, the current experimental bounds can be used to constrain model parameters.
FCNC interaction with SM4 has been studied for a long time\cite{Ref:SM4q1} and can
also help for solving some of open questions in FCNC quark interaction\cite{Ref:SM4q2}.
Leptonic FCNC effects, including $\mu-e$ conversion, in SM4 have also been studied extensively\cite{Ref:SM4LFV1,Ref:SM4LFV2,Ref:SM4LFV3,Ref:SM4LFV4}.
In our work we carry out a more detailed systematic study by taking account of all available $\mu -e$ conversion experimental results.
We compare constraints obtained from $\mu - e$ conversion using experimental bounds on various nuclei with
those from $\mu \to e \gamma$ and $\mu \to e\bar e e$. We find that the current bound from $\mu - e$ conversion with Au puts the most
stringent constraints on model parameters. The relevant flavor changing
parameter $\lambda_{\mu e} = V^*_{\mu 4}V_{e4}$ is constrained to be less than $1.6\times 10^{-5}$ for $m_{\nu_4}$ larger than $100$ GeV. Future improved experiments on $\mu - e$ conversion, $\mu \to e\gamma$ and $\mu \to e \bar e e$ will further constrain model parameters.

\section{The FCNC interactions for $\mu -e$ conversion in SM4 }

The minimal set of particle contents in the SM is the gauge bosons of the
gauge group $SU(3)_C\times SU(2)_L\times U(1)_Y$ with couplings $g_s$, $g$ and $g'$, the Higgs doublet $\Phi: (1,2,+1)$ with the component fields and vacuum expectation value (vev) $v = \langle h\rangle$
given by $\Phi = (\omega^+, (v+ h + i z)/\sqrt{2})^T$, the three generations of left and right
handed up and down quarks, $Q_L: (3,2,1/3)$, $U_R: (3,1,4/3)$, $D_R: (3,1,-2/3)$, and the three generations of left handed leptons $L_L: (1,2,-1)$ and
right handed charged leptons, $E_R: (1,1,-2)$. We use the hypercharge normalization that $Q = I_3 + Y/2$. Depending on the nature of the neutrinos, there may be right handed neutrinos,
$N_R: (1,1,0)$ which may pair up with the left handed neutrinos to have Dirac or by themselves to have Majorana masses.

For a model with Dirac neutrino mass only, the charged lepton and neutrino masses are generated by the Yukawa interactions
\begin{eqnarray}
L = - \bar L_L Y_E \Phi E_R - \bar L_L Y_N \tilde \Phi N_R + h.c.
\label{dirac-neu}
\end{eqnarray}
where $\tilde \Phi = i\sigma_2 \Phi^*$,
$E_R = (e, \mu, \tau, ...)_R^T$, $N_R = (\nu_e, \nu_\mu, \nu_\tau, ...)_R^T$ for $n$ generations of leptons. $Y_E$ and $Y_\nu$ are $n\times n$ matrices.

After the Higgs boson develops a non-zero vev, the leptons will obtain their masses with the
mass matrices for charged leptons and neutrinos which are given by
\begin{eqnarray}
M_E = Y_E \frac{v}{\sqrt2}\;,\;\;M_\nu = Y_N \frac{v}{\sqrt2}\;.
\end{eqnarray}

In the basis where the mass matrices are all diagonal, the flavor changing interactions involving leptons are
contained in the PMNS mixing matrix $U=(V_{ij})$ defined by
\begin{eqnarray}
L = - {g\over \sqrt{2}} \bar E_L \gamma^\mu U \nu_L W^-_\mu + h.c.
\label{sm}
\end{eqnarray}
where $E_L = (e, \mu, \tau, ...)_L^T$, $\nu_L = (\nu_1, \nu_2, \nu_3, ...)_L^T$, and $U$ is a $n\times n$ unitary matrix for $n$ generations.

For SM3, $U$ can be parameterized as\cite{Ref:PDG}
\begin{eqnarray}
U = \left (\begin{array}{ccc}
c_{12}c_{13}& s_{12}c_{13}&s_{13}e^{-i\delta}\\
-s_{12}c_{23} - c_{12}s_{23}s_{13}e^{i\delta}& c_{12}c_{23}-s_{12}s_{23}s_{13}e^{i\delta}&s_{23}c_{13}\\
s_{12}s_{23}-c_{12}c_{23}s_{13}e^{i\delta}& -c_{12}s_{23}-s_{12}c_{23}s_{13} e^{i\delta}&c_{23}c_{13}\end{array}
\right )\;,
\end{eqnarray}
where $c_{ij} = \cos\theta_{ij}$ and $s_{ij} = \sin\theta_{ij}$ with $\theta_{ij}$ being rotation angles and $\delta$ being the CP violating phase.

The charged lepton masses have been measured with great precision. But the neutrino masses are not. The absolute values for neutrino mass scale, although not known, are constrained to be small
($m_\nu < 2.3$ eV by tritium beta decay experiment\cite{Ref:mnu-beta}, and $\sum m_i < 0.61$ eV by cosmological observations\cite{Ref:mnu-cosmo}), and the mass differences are known from neutrino oscillation data. The mixing angles are also constrained by various experimental data in particular from neutrino oscillation data.
We list these constraints on neutrino mixing and masses in the following
\begin{eqnarray}
&&\sin^2(2\theta_{12})=0.87\pm0.23\;\text{\cite{Ref:th12}},\;
\sin^2(2\theta_{23})>0.92\;\text{\cite{Ref:th23}},\;
\sin^2(2\theta_{13})<0.15\;\;\text{CL}=90\%\text{\cite{Ref:th13}},\nonumber\\
&&\Delta m_{21}^2 = (7.59\pm0.20) \times 10^{-5}\;\text{\cite{Ref:th12}},\;
|\Delta m_{31}^2| = (2.43\pm0.13) \times 10^{-3} \text{eV}^2\text{\cite{Ref:dm31}},
\label{ct-mixing}
\end{eqnarray}
where $\Delta m^2_{ij} = m_i^2-m_j^2$, and $m_i(i=1$--$3)$ are the mass eigenvalues of neutrinos.
Recently T2K has obtained new evidence\cite{Ref:t2k} showing that the angle $\theta_{13}$ may indeed be non-zero with the 90\% c.l. allowed region $0.03(0.04)\sim 0.28(0.34)$ for $\sin^2(2\theta_{13})$ for normal (inverted) neutrino mass hierarchy.

In this model FCNC processes do not occur at the tree level, but it can happen at one loop level. Due to GIM mechanism of the model for lepton sector, amplitudes for any leptonic FCNC processes with charged leptons are proportional to $\Delta m_{ij}^2/m^2_W$.
Since squared differences of neutrino masses are constrained as in above,
all FCNC processes involve charged leptons including $\mu - e$ conversion, are very small which are well below experimental sensitivity. To have observable FCNC effects, further extensions are needed. A simple way which can overcome the problem of small $\Delta m_{ij}^2/m^2_W$
ratio is to have a heavy neutrino in the model. This can be achieved for a model with four generations (SM4).

The existence of a fourth generation of quarks and leptons is not ruled out although there are severe constraints.
The heavy neutrino must be heavier than $m_Z^{}/2$ such that the precisely measured $Z$ decay width would not be upset. The LEP II data constraint the mass to be even larger with $m_{\nu_4}>90.3$\text{GeV} at $95\%\text{CL}$~\cite{Ref:mnu4-limit}.
The constraint from electroweak precision data can also be evaded by imposing appropriate mass splitting between fourth generations\cite{Ref:s-param}.
Perturbative unitarity bound on a heavy fourth generation
Yukawa coupling constant has been considered\cite{Ref:UB-fermions}.
Critical masses determined by perturbativity for charged leptons and neutrinos are about $1.2$ TeV\cite{Ref:UB-fermions}. The fourth generation neutrinos, if exists, with a large mass in the range about $100$ GeV to the unitarity bound are not ruled out. With the addition of the fourth generation, the PMNS matrix $U$ will become a $4\times 4$ unitary matrix. If mixing matrix elements in $V_{i4}$ are not extremely small, larger FCNC effects can be generated at one loop level. From the neutrino oscillation data, the constraints on the mixing of the fourth generation with the first three light neutrinos are indeed allowed to be not extremely small as the large error bars in eq.\eqref{ct-mixing} indicate. There is a chance for large FCNC effects in the lepton sector.

The strength of muonic FCNC processes is governed by powers of $\lambda_{\mu e} = V^*_{\mu 4} V_{e4}^{}$. Loop induced FCNC interactions relevant to $\mu -e $ conversion can be obtained by standard loop calculations. We present the one loop effective Lagrangian in two terms in the following, in the limit of neglecting terms suppressed by light neutrino mass squared divided by the $W$ mass squared,
\begin{eqnarray}
&&L(\mu \to e \gamma) = -{4 G_F\over \sqrt{2}}{ e \over 32 \pi^2 } \lambda_{\mu e} G_2(x_4)\bar e \sigma^{\mu\nu} (m_\mu R + m_e L)\mu F_{\mu\nu}\;,\nonumber\\
&&L(\mu \to e  + q\bar q)=-{G_F \over \sqrt{2}} {e^2 \over 4 \pi^2 }\lambda_{\mu e} \left [ V_u(x_4) \bar u \gamma_\mu u+ V_d(x_4) \bar d \gamma_\mu d \right .\nonumber\\
&&\;\;\;\;\;\;\;\;\;\;\;\;\;\;\;\;\;\;\;\;\;\;\;\left .+ A_u(x_4) \bar u \gamma_\mu\gamma_5 u+ A_d(x_4) \bar d \gamma_\mu\gamma_5 d\right ]\bar e \gamma^\mu L \mu \;,
\label{basic1}
\end{eqnarray}
where $\sigma_{\mu\nu} = i[\gamma_\mu, \gamma_\nu]/2$, $R(L) = (1\pm \gamma_5)/2$, $x_4 = m^2_{\nu_4}/m^2_W$, and
\begin{eqnarray}
&&V_q(x) = -Q^q G_1(x) + {1\over 2 s^2_W} (I_3^q-Q^q s^2_W)G_Z(x) - {1\over 4 s^2_W} G_B^q(x)\;,\nonumber\\
&&A_q(x) = -{1\over s^2_W} I_3 G_Z(x) + {1\over 4 s^2_W} G_B^q(x)\;,\nonumber\\
&&G_1(x) = {x(12+x-7 x^2)\over 12(x-1)^3} - {x^2(12-10x+x^2)\over 6(x-1)^4}\ln x\;,\nonumber\\
&&G_2(x) = {x(1-5x-2x^2)\over 4 (x-1)^3} + {3x^3\over 2 (x-1)^4}\ln x\;,\nonumber\\
&&G_Z(x) = {x(x^2-7x+6)\over 4(x-1)^2} + {x(2+3x)\over 4 (x-1)^2}\ln x\;,\nonumber\\
&&G_B(x) = {x\over x-1} - {x\over (x-1)^2}\ln x\;,\;\;
G^u_B(x) = 4G_B(x)\;,\;\;G^{d}_B(x) = G_B(x)\;.
\end{eqnarray}
In the above, $Q^q$ is the electric charge of the quark $q$.

We also present the effective Lagrangian for $\mu \to e\bar e e$ in the following which will be used when we compare constraints on parameters,
\begin{eqnarray}
L(\mu \to e \bar e e) &=& -{G_F \over \sqrt{2}}{e^2\over 4 \pi^2 } \lambda_{\mu e} \Bigl[ G_2(x_4)\,
\bar e \gamma_\mu e\, {q_\nu\over q^2}\, \bar e i\sigma^{\mu\nu}(m_\mu R + m_e L) \mu  \nonumber\\
 &+&  \bar e \gamma^\mu \left ( a_L(x_4) L+ a_R(x_4) R \right ) e\, \bar e \gamma_\mu L \mu \Bigr] \;.
 \label{basic2}
\end{eqnarray}
where
\begin{eqnarray}
a_L(x) = G_1(x) + {1\over s^2_W}\Bigl(-{1\over 2} + s^2_W\Bigr) G_Z(x) - {1\over 2 s^2_W} G_B(x)\;,\;\;\;\;a_R(x) = G_1(x) + G_Z(x)\;.
\end{eqnarray}

The term $L(\mu \to e\gamma)$ is generated by the electromagnetic penguin.
The loop generates the function $G_2(x)$ for dipole operator. For the terms $L(\mu \to e + q\bar q)$ and $L(\mu \to e \bar e e)$,
there are several contributions. Connecting the photon in the dipole operator, this generates the terms proportional to $G_2(x)$. The electromagnetic penguin also generates
a charge radius term. With the photon connected to a quark or an electron, the terms proportional to $G_1(x)$ are generated. Replacing the photon by a $Z$ boson, terms proportional to $G_Z(x)$ are generated. Finally, the box diagram produce the terms proportional to $G_B(x)$.
We note that in the large $x$ limit the function $G_Z(x)$ linearly increases, while others are not.
The results above agree with those in Ref.\cite{Ref:SM4LFV3}.

\section{$\mu \to e$ conversion constraints on model parameters}

To obtain $\mu \to e$ conversion rates from the FCNC interaction discussed in the previous section, one needs to convert quarks into hadrons which depend on hadronic matrix element calculations.
Theoretical efforts have been made by several group to calculated these matrix elements\cite{Ref:Mu2Ecv3,Ref:Mu2Ecv4,Ref:Mu2Ecv1}. We will use the results in Ref.\cite{Ref:Mu2Ecv1} for a consistent calculation. In this framework there are several quark level contributions which are parameterized as the following effective Lagrangian
\begin{eqnarray}
L_{eff} &=& -{4 G_F\over \sqrt{2}} \left [m_\mu \bar e \sigma^{\mu\nu} (A_R R + A_L L) \mu F_{\mu\nu}  + h.c.\right ]\nonumber\\
& - & {G_F\over \sqrt{2}} \left [ \bar e (g_{LS(q)}  R  + g_{RS(q)} L )\mu\, \bar q q  + \bar e (g_{LP(q)}  R  + g_{RP(q)} L )\mu\, \bar q \gamma_5 q+ h.c.\right ]\nonumber\\
& - & {G_F\over \sqrt{2}} \left [ \bar e (g_{LV(q)} \gamma^\mu L  + g_{RV(q)} \gamma^\mu R )\mu\, \bar q\gamma_\mu  q  + \bar e (g_{LA(q)}  \gamma^\mu L  + g_{RA(q)} \gamma^\mu R )\mu\, \bar q \gamma_\mu \gamma_5 q + h.c. \right ]\nonumber\\
&-& {G_F\over \sqrt{2}} \left [ {1\over 2}\bar e (g_{LT(q)} \sigma^{\mu\nu} R  + g_{RT(q)} \sigma^{\mu\nu} L )\mu\, \bar q\sigma_{\mu\nu}  q + h.c.\right ]\;.
\end{eqnarray}

In SM4, at the one loop level, we have, $g_{(L,R)(S,P,T)(q)} = 0$, $g_{RV(q)}=0$, and
\begin{eqnarray}
&&A_R =  {e\over 32 \pi^2} \lambda_{\mu e} G_2(x_4) \;,\;\; A_L = {m_e \over m_\mu} A_R\;,\;\;g_{LV(q)} = {e^2\over 4 \pi^2}\lambda_{\mu e} V_q\;.
\label{arl}
\end{eqnarray}
The contribution from $A_L$ can be neglected compared with that from $A_R$. The contribution from $g_{L(R)A}$ terms proportional to $A_q$ in eq.(\ref{basic1}) is suppressed because the fraction of the coherent process is generally larger than that of the incoherent one approximately by a factor of the mass number of the target nuclei\cite{Ref:Mu2Ecv1}.

The quantity $B^A_{\mu \to e}$ measuring the leptonic FCNC effect in $\mu\to e$ conversion to the leading order in SM4 is proportional to
\begin{eqnarray}
|A_R D + \tilde g^{(p)}_{LV} V^{(p)} + \tilde g^{(n)}_{LV} V^{(n)}|^2\;,
\end{eqnarray}
where
\begin{eqnarray}
\tilde g^{(p)}_{LV} = 2 g_{LV(u)}+g_{LV(d)}\;,\;\;\tilde g^{(n)}_{LV} = g_{LV(u)}+ 2 g_{LV(d)}\;.
\end{eqnarray}
Combining these, we obtain
\begin{eqnarray}
{B^A_{\mu\to e}\over B(\mu \to e\gamma)} = R^0_{\mu\to e}(A) \left | 1 + {\tilde g^{(p)}_{LV} V^{(p)}(A)\over A_R D(A)} + {\tilde g^{(n)}_{LV} V^{(n)}(A)\over A_R D(A)}\right |^2\;,
\end{eqnarray}
where
\begin{eqnarray}
R^0_{\mu\to e}(A) = {G^2_F m^5_\mu \over 192 \pi^2 \Gamma^A_{capt}}|D(A)|^2\;.
\end{eqnarray}
The relevant parameters are listed  in Table \ref{Tab:mu-e}, which are
evaluated using method I in Ref.\cite{Ref:Mu2Ecv1}.
\begin{table}[tb]
\begin{tabular}{l||l|l|l|l}
A & $D$(A) & $V^{(p)}$(A) & $V^{(n))}$(A) & $R^0_{\mu\to e}$(A) \\
\hline \hline
${}^{27}_{13}$Al  & 0.0362 & 0.0161 & 0.0173 & 0.0026 \\
${}^{32}_{16}$S   & 0.0524 & 0.0236 & 0.0236 & 0.0028 \\
${}^{48}_{22}$Ti  & 0.0864 & 0.0396 & 0.0468 & 0.0041 \\
${}^{197}_{79}$Au & 0.189  & 0.0974 & 0.146  & 0.0039 \\
${}^{208}_{82}$Pb & 0.161  & 0.0834 & 0.128  & 0.0027
\end{tabular}
\caption{The relevant parameters for $\mu - e$ conversion processes, which are evaluated
by using method I in Ref.\cite{Ref:Mu2Ecv1}.}
\label{Tab:mu-e}
\end{table}

\begin{figure}[tb]
\includegraphics[width=9cm]{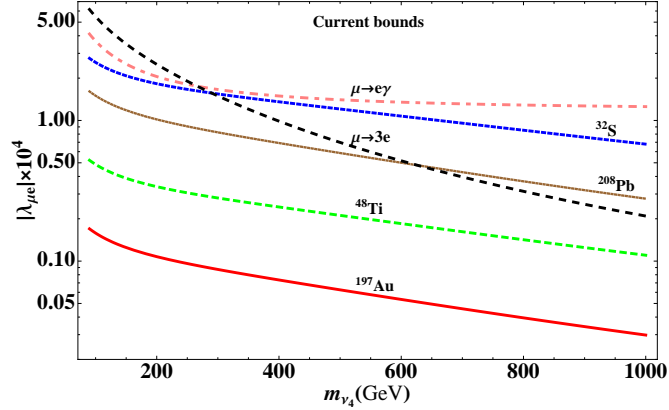}
\caption{Constraints on $|\lambda_{\mu e}|$ as a function of neutrino mass $m_{\nu_4}$ from current experimental bounds on $\mu - e$ conversion, $\mu \to e\gamma$ and $\mu \to e \bar e e$.}\label{mu-e-bound}
\end{figure}

In Fig.\ref{mu-e-bound} we show constraints on $\lambda_{\mu e}$ as a function of the fourth generation neutrino mass
$m_{\nu_4}$ using experimental upper limits on $\mu -e $ conversion for Au, S, Ti and Pb nuclei. The most stringent constraint comes from Au.
We see that $\lambda_{\mu e}$ is constrained to be less than $1.6\times 10^{-5}$ for $m_{\nu_i}$ larger than $100$ GeV. Such a small value for $\lambda_{\mu e}$ is well within the error bars allowed ranges from neutrino oscillation data. To see whether $\mu -e$ conversion provides the strongest constrains on the model parameters, one needs to check with other leptonic FCNC processes. In the next section we compare the $\mu -e$ conversion constraints with those obtained from other two well known leptonic FCNC processes $\mu \to e\gamma$ and $\mu \to e\bar e e$.

\section{Comparisons with constraints from $\mu \to e \gamma$ and $\mu \to e\bar e e$}

$\mu \to e \gamma$ and $\mu \to e\bar e e$ are two well known processes which provide stringent constraints on leptonic FCNC interactions.
Experimental measurement for $\mu \to e\gamma$ is not an easy task. Nevertheless, several groups have made efforts in obtaining limit on this decay.
No positive result on the observation of this decay has been reported. The
best upper bound at the $90\%$CL, set more than $10$ years ago, is $B(\mu \to e\gamma) < 1.2\times 10^{-11}$\cite{Ref:MEGA}.
Recently MEG collaboration has reported their preliminary result on the upper limit with $B(\mu \to e\gamma) < 1.5\times 10^{-11}$  based on their 2009 physics data collection\cite{Ref:MEG09}. This bound is slightly weaker than the best upper bound.
It is, however, intriguing to note that they did have a few events well separated from backgrounds in their photon and electron energy cuts plot. However, due to low statistics, no positive claims can be made.
MEG experiment has potential to improve the sensitivity by two orders of magnitudes, $10^{-13}$\cite{Ref:MEG-pot}.
Experimental studies of $\mu \to e\bar e e$ has put a stringent limit on the branching ratio. The current best limit is $B(\mu \to e \bar e e) < 1.0\times 10^{-12}$\cite{Ref:SINDRUM-m3e}.

Using the effective Lagrangian $L(\mu \to e\gamma)$ in eq. \eqref{basic1}, one can obtain the branching ratio $B(\mu \to e \gamma) = \Gamma(\mu \to e \gamma) /\Gamma(\mu \to e \nu \bar \nu)$.
Neglecting $m_e$, one obtains
\begin{eqnarray}
B(\mu \to e \gamma) = 384\pi^2 (|A_L|^2 + |A_R|^2)\;,
\end{eqnarray}
where $A_{R,L}$ are defined in eq.\eqref{arl}.
If the mass $m_e$ is kept, one should divide, in the above expression, a phase factor $I(x) = 1 -8x+8x^3 -x^4 -12 x^2\ln x$ with $x = m^2_e/m^2_\mu$.
Radiative corrections from QED also modify the above expression by dividing a factor $1+\delta_{QED}$ with $\delta_{QED} = (\alpha/2\pi)(25/4 - \pi^2)$.

The branching ratio for $\mu \to e\bar e e$ can be deduced from $L(\mu \to e \bar e e)$ in eq.\eqref{basic2}. We have
\begin{eqnarray}
B(\mu \to e\bar e e) &=& {\alpha^2\over 16 \pi^2} |\lambda_{\mu e}|^2\left [ a_R^2(x_4) + 2 a_L^2(x_4)\right . \nonumber\\
&-&\left . 4 G_2(x_4)(a_R(x_4)+2a_L(x_4)) + 4 G^2_2(x_4)\Bigl(4 \ln{m_\mu\over  m_e} - {11\over 2}\Bigr)\right ]\;.
\end{eqnarray}

The constraints on the parameter $\lambda_{\mu e}$ as functions of $m_{\nu_4}$ for bounds from $\mu \to e \gamma$ and $\mu \to e\bar e e$ are also shown in Fig.\ref{mu-e-bound}. The constraint obtained from $\mu \to e\bar e e$ is weaker for neutrino mass $m_{\nu4}$ less than $255$ GeV than that from $\mu \to e\gamma$, but becomes stronger for larger mass for the current experimental bounds. This property is mainly caused by $Z$-penguin contributions. However, the constraints from $\mu -e$ conversion on Ti and Au are better than those from $\mu \to e \gamma$ and $\mu \to e\bar e e$. The best constraint comes from $\mu -e$ conversion on Au nuclei.

\begin{figure}[tb]
\includegraphics[width=8.5cm]{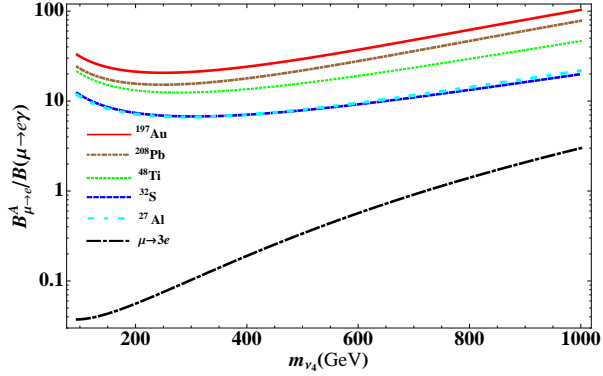}
\caption{Ratios of $B^A_{(\mu -e)}/B(\mu \to e\gamma)$ and $B(\mu \to e \bar e e)/B(\mu \to e\gamma)$ as functions of $m_{\nu_4}$. }\label{Ratio}
\end{figure}

In Fig.\ref{Ratio} we show the ratios of $B^A_{(\mu -e)}/B(\mu \to e\gamma)$ and $B(\mu \to e \bar e e)/B(\mu \to e\gamma)$ as functions of $m_{\nu_4}$. For a given fourth generation neutrino mass, the ratios are fixed independent the value of $\lambda_{\mu e}$. We see that in the region with $m_{\nu_4}$ larger than $100$ GeV, the ratios for $\mu - e$ conversion of various nuclei
are larger than one implying that if the experimental values for the $\mu -e $ conversion rate and $\mu \to e\gamma$ branching ratio are similar, the constraint from $\mu -e $ conversion will give stronger constraints on the parameter $\lambda_{\mu e}$ which explains the fact that the current experimental bounds on $\mu - e$ conversion from Ti and Au provide stronger constraints.
When $m_{\nu_4}$ becomes larger ratios become larger due to the fact that the $Z$ penguin whose contribution increases with $m^2_{\nu_4}$ for $\mu - e$ conversion and $\mu \to e \bar e e$, but $\mu \to e \gamma$ increases only logarithmically in large $m_{\nu_4}$ limit.

\begin{figure}[tb]
\includegraphics[width=9cm]{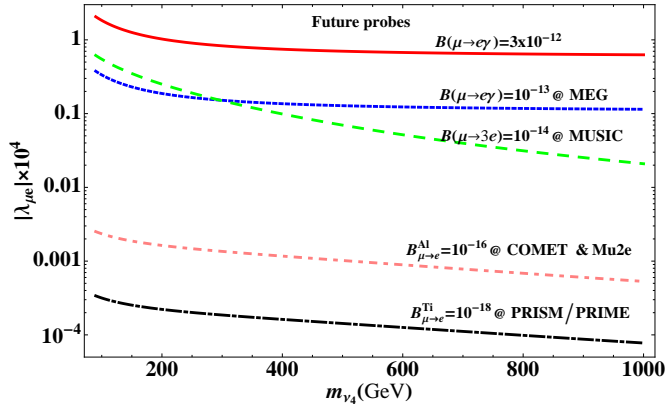}
\caption{Constraints on 4th-generation mixing coupling $|\lambda_{\mu e}|$ as a function of neutrino mass $m_{\nu_4}$ from several future experimental sensitivities on $\mu - e$ conversion, $\mu \to e\gamma$ and $\mu \to e \bar e e$.}\label{future}
\end{figure}

If one takes the well separated events of the MEG data as $\mu \to e \gamma$ events, the branching ratio implied would be of order $10^{-12}$, we will take an assumed branching ratio of $3\times 10^{-12}$ to show the implication on the parameters required to produce it. We show the results in Fig.\ref{future}. We see that such a possibility is ruled out by $\mu -e$ conversion on Au nuclei in this model (see also Fig.\ref{mu-e-bound}). In Fig.\ref{future}, we also plot possible constraints on parameters using several projected experimental sensitivities on the $\mu -e$ conversion, $\mu \to e\gamma$ and $\mu\to e \bar e e$. We see that the improved experiment MEG on $\mu \to e \gamma$  and MuSIC\cite{Ref:MuSIC} on $\mu\to e\bar e e$ can obtain better constraints than the current Au bound. Future experiments Mu2E\cite{Ref:Mu2E}/COMET\cite{Ref:COMET} and PRISM\cite{Ref:PRISM} for $\mu -e $ conversion using Al and Ti, respectively, can obtain much better constraints.

\section{Conclusions}

We have studied $\mu - e$ conversion in the SM with four generations. A large mass $m_{\nu_4}$ for the fourth generation neutrino can enhance the conversion rate by orders of magnitude. We have compared constraints obtained from $\mu - e$ conversion using experimental bounds on various nuclei with
those from $\mu \to e \gamma$ and $\mu \to e\bar e e$. We found that the current bound from $\mu - e$ conversion with Au puts the most stringent constraint in this model. The relevant flavor
changing parameter $\lambda_{\mu e} = V^*_{\mu 4}V_{e4}$ is constrained to be less than $1.6\times 10^{-5}$ for $m_{\nu_4}$ larger than 100 GeV.
The goal of MEG experiment is to have a sensitivity of $10^{-13}$ for $B(\mu\to e\gamma)$, which can compete that of $\mu - e$ conversion in Au in this model. Therefore, the model prediction can be tested
by combining these coming data.
There are several proposed experiment for $\mu - e$ conversion, Mu2E\cite{Ref:Mu2E}, COMET\cite{Ref:COMET} and PRISM\cite{Ref:PRISM}, which will tell us detailed information of lepton flavor mixing.
In addition, successful running and energy upgrade of Large Hadron Collider will search for sequential fourth generation fermions. The fourth generation model will be explored directly and indirectly by future experiments in great detail.

{\bf Acknowledgment}:

We thank Dr. Goto and Dr. Nomura for useful discussions. This work was partially supported by USA DOE grant number DE-FG02-96ER40969, Grant-in-Aid for Scientific Research from the Ministry of Education, Science and Culture of Japan, \#20540282 and \#21104004, NSC and NCTS of ROC, and SJTU 985 grant of China.

\end{document}